# STOCHASTIC THEORY OF QUANTUM MECHANICS

Maurice GODART,
Email = maurice.godart@skynet.be

**0   Introduction**

The questions on the underlying reality behind the laws and procedures of quantum mechanics have been a battlefield over the entire period since its initial development, and no truce is yet in sight despite an extensive literature that refers to, discusses and criticizes the orthodox Copenhagen interpretation.

On the one hand its defenders claim that the validity of the theory is justified by the many successes of its predictions and applications. In our opinion this is not at all a convincing argument, because many examples mainly from the past show that a solidly founded theory is not always necessary to arrive at spectacularly useful results, as attested by the fact that empiricism has always been a powerful method ([1]).

On the other hand its opponents wonder if it was absolutely necessary to introduce the notions of duality, complementarity and contraction of the wave packet. Moreover the essential role played by the process of measurement is a hotly contested point, because it is in opposition with the idea that a physical theory must provide an objective description of the phenomena observed.

The inspiration for this work came from two important publications. The first of them is a book written by Landé ([2]). It is an incisive critique of the Copenhagen interpretation of quantum mechanics, but certainly not of its results. Landé significantly claims that "...if Bohr and Heisenberg are severely criticized in this [book] for their subjectivist approach, their language refinement, and their occasional deviation from logic and consistency, it will hardly be necessary to add that my admiration for their achievements in theoretical physics is as great as ever. But I must confess that, during many years of trying to imbue my students with the Copenhagen Spirit, I felt more and more like the Devil's Advocate, suffering from an ever-growing intellectual distress".

The book starts with a formidable thunderbolt in a cloudless blue sky. Landé reports the existence of a purely mechanical explanation for the electron diffraction experiment due to Duane ([3]). The resulting rule states that the angles of reflection of particles (or photons) arriving with momentum p on a crystal with parallel planes spaced at distance L are selected in agreement with the rule:

$\Delta p = n\, h/L$

---

[1] Remember that the Damascus steel was highly praised at the time of the Crusades and thus had been discovered and was produced without the help of modern metallurgy and crystallography. It went forgotten mysteriously and only recently did experts in those domains claim to have rediscovered this lost secret.

[2] Landé, A.
New foundations of quantum mechanics.
Cambridge University Press, 1965, xii+171pp.

[3] Duane, W.
The transfer in quanta of radiation momentum to matter.
Proc. Nat. Acad. Sci. Wash., 9, 1923, 158-164.

Duane rule is as legitimate as the two familiar old quantum rules for the energy E and for the angular momentum M, namely Planck rule $\Delta E = h/T$ for bodies characterized by a periodicity in time with period T (harmonic oscillator) and Sommerfeld-Wilson rule $\Delta M = h/2\pi$ for bodies periodic with respect to rotations through $2\pi$ (any body). The question thus arises to know whether duality would have been invented if Duane rule had been taken into account in the pioneering years of quantum mechanics. Let us also mention that the results of the famous two slits experiment can be explained mechanically by a further development of Duane ideas applied to the theory of diffraction ([1]).

We must however confess that Landé raises many pertinent questions, but does not offer satisfactory answers. Quite fortunately a global solution was suggested in a second publication written by Nelson ([2]). who presents a plausible interpretation of quantum theory

At the risk of contributing to the confusion reigning supreme in this domain, this paper will defend this interpretation. In our opinion the solution to the problem of interpretation consists in providing some form of plausible picture. This can be achieved by purely mathematical solutions as offered by the stochastic theory of quantum mechanics. As its name suggests it relies heavily upon the theory of stochastic processes, with its definitions, theorems and specific vocabulary as well. Its main hypothesis states indeed that the classical trajectories of the particles are identical to the sample functions of a diffusion Markov process, whose conditional probability density is proposed as a substitute for the wave function.

The present work can be seen as a presentation to other interested people of attempts to understand the theory of Nelson, to explain it to anybody so to say, and to extend it as far as possible in a coherent view of the quantum world.

This work pertains clearly to the domain of mathematical physics. A compromise inconsistent as most compromises are had to be made because this work on the one hand calls for a rigorous treatment guaranteeing solid foundations, but on the other hand is threatened with the charge of being absurdly overloaded with extraneous higher mathematics. This explains the fact that many proofs are (over-)simplified or even completely left aside. On the contrary others can appear to be too much detailed. Another (personal) reason is that we hate expressions like "it is well-known that" and we tried therefore to make this work self contained as far as possible. The complete text with its 190 pages is much too long to be published here and we must be satisfied by presenting here a detailed table of contents of it.

# 1    Probability concepts

This chapter presents the basic concepts about probability, conditional probability, probability densities, conditional probability densities, mathematical expectations and conditional mathematical expectations applied to the random variables of stochastic processes that are absolutely necessary for the reading and understanding of the complete text. The readers who are already experts in those domains can safely go to the next chapter.

---

[1] Epstein, P. S. - Ehrenfest, P.
The quantum theory of the Fraunhofer diffraction.
Proc. Nat. Acad. Sci. Wash., 10, 1924, 133-139
[2] Nelson, E.
Derivation of the Schrödinger equation from Newtonian mechanics.
Phys. Rev., 150, 1966, 1079-1085.

From the point of view of the expert, probability is simply a branch of measure theory, with its own special emphasis and fields of application. There is a familiar inclination on the part of those who possess unusual and arduously acquired expertise to exaggerate its remoteness from anything the rest of us know, but we must warn you that there is perhaps no mathematical subject that shares with probability the characteristic features that on the one hand many of its most elementary theorems are based on deep mathematics, and on the other hand that many of its advanced theorems are often considered as "intuitively obvious" by many people who are ready to accept them without proofs.

We present the probability concepts first very briefly in the general case where a distribution function $P\{a \leq x < b\}$ gives the probability to find the value of the random variable x within the indicated interval, and then with more details in the simpler case where there exists a probability density function $p(x)$ such that we can write:

$$P\{a \leq x < b\} = \int_a^b p(x) dx$$

The (basic) concept of mathematical expectation is then defined. All the points presented up to here are easily generalized to the case of several random variables.

The (less basic) concepts of conditional probability and conditional mathematical expectation are then presented, first briefly in the general context where a distribution function $P\{a \leq x < b | M\}$ gives the probability to find the value of the random variable x within the indicated interval under the additional circumstance that the condition M must be fulfilled, and then with more details in the simpler context where there exists a probability density function $p(x|M)$ such that we can write:

$$P\{a \leq x < b | M\} = \int_a^b p(x|M) dx$$

We then introduce the notation $p(x_1 | x_0)$ in the special case important for the future where we are interested by the conditional density probability of a random variable $x_1$ under the condition that another random $x_0$ variable takes on imposed values. We show that in the case where the probability densities $p(x_0, x_1)$ and $p(x_0)$ exist we have:

$$p(x_1 | x_0) = p(x_0, x_1) / p(x_0)$$

The main properties of the normal or conditional probability densities and mathematical expectations are proved and generalized to the case of several random variables.

The important special case of so-called independent random variables is finally examined in all details necessary for the future developments.

## 2 Stochastic processes.

The subject of the first chapter can seem futile to the experts. We are however convinced that the theory of probability as taught at the University often serves exclusively to introduce statistics with its confidence intervals, Bayesian rules, Gaussian distribution and all the like. The theory of stochastic processes if ever addressed is severely restricted to the domain of the Poisson distributions and of the Markov chains.

This chapter defines roughly a stochastic process as being a random variable that depends on a parameter t usually identified with the time. The special case of a Wiener process is examined in details as an exercise and because of its importance for the future.

General Markov processes are defined as those obeying the Chapman-Kolmogorov relation, and the diffusion Markov processes are then introduced as those special cases characterized by the existence of a drift velocity v and a diffusion tensor w. It is proved that the probability density $p(x,t)$ satisfies a partial differential equation called the Fokker-Planck equation and that the conditional probability density $p(x_1,t_1 | x_0,t_0)$ verifies two partial differential equations, one for each pair of variables x and t, named after Kolmogorov who was the first to discover them. The natural initial condition:

$$p(\bar{x},t_0 | \bar{x}_0,t_0) = \delta(x - x_0)$$

is proposed for it. This is easily extended from the case of a scalar stochastic process to the case of a vector stochastic process.

It is shown that a diffusion Markov process can be equivalently defined as the solution of a particular system of difference or differential stochastic equations. This point of view has its own merit because it sheds light on curious properties of the trajectories and it leads to uncertainty relations. At this point, mean velocities and stochastic derivatives with respect to t of a stochastic process are introduced as a replacement for the classical velocities and derivatives.

A Markov process can be defined otherwise by the fact that its conditional probability densities must satisfy the relations:

$$p(\bar{x}_n,t_n | \bar{x}_1,t_1; \cdots; \bar{x}_{n-1},t_{n-1}) = p(\bar{x}_n,t_n | \bar{x}_{n-1},t_{n-1})$$

for any value of n and for any increasing sequence $t_1 < t_2 < \cdots < t_n$ of values for the parameter t. This condition that makes play a role to the direction of time thus seems to introduce an element of irreversibility but this is far from true because a Markov process also satisfies also the relations:

$$p(\bar{x}_1,t_1 | \bar{x}_2,t_2; \cdots; \bar{x}_n,t_n) = p(\bar{x}_1,t_1 | \bar{x}_2,t_2)$$

for the same value of n and for the same sequence of values for t.

This property is expressed by saying that the process is reversible, and this implies that the definitions of drift velocities, diffusion tensors, and that the partial differential equations of Fokker-Planck and of Kolmogorov can be written twice, in a forward version generally using indices + when the direction of time is from the present to the future and in a backward version generally using indices - when the direction of time is from the present to the past.

So for example we must consider simultaneously the two definitions of the drift velocities:

$$\int_{-\infty}^{\infty} (x_1^i - x_0^i) p(x_1,t_1 | x_0,t_0) dx_1 = v_{\pm}^i(x_0,t_0)(t_1 - t_0)$$

the two diffusion tensors:

$$\int_{-\infty}^{\infty} (x_1^i - x_0^i)(x_1^j - x_0^j) p(x_1, t_1 \mid x_0, t_0) dx_1 = 2 w_+^{ij}(x_0, t_0) |t_1 - t_0|$$

the two Fokker-Planck equations:

$$\frac{\partial p(\overline{x}, t)}{\partial t} + \sum_{i=1}^{n} \frac{\partial}{\partial x^i} \left[ v_\pm^i(\overline{x}(t), t) p(\overline{x}, t) \right]$$

$$\mp \sum_{i=1}^{n} \sum_{j=1}^{n} \frac{\partial^2}{\partial x^i \partial x^j} \left[ w_\pm^{ij}(\overline{x}(t), t) p(\overline{x}, t) \right] = 0$$

the two first Kolmogorov equations:

$$\frac{\partial p(\overline{x}_1, t_1 \mid \overline{x}_0, t_0)}{\partial t_0} + \sum_{i=1}^{n} v_\pm^i(\overline{x}(t_0), t_0) \frac{\partial p(\overline{x}_1, t_1 \mid \overline{x}_0, t_0)}{\partial x_0^i}$$

$$\pm \sum_{i=1}^{n} \sum_{j=1}^{n} w_\pm^{ij}(x(t_0), t_0) \frac{\partial^2 p(\overline{x}_1, t_1 \mid \overline{x}_0, t_0)}{\partial x_0^i \partial x_0^j} = 0$$

and the two second Kolmogorov equations:

$$\frac{\partial p(\overline{x}_1, t_1 \mid \overline{x}_0, t_0)}{\partial t_1} + \sum_{i=1}^{n} \frac{\partial}{\partial x_1^i} \left[ v_\pm^i(\overline{x}(t_1), t_1) p(\overline{x}_1, t_1 \mid \overline{x}_0, t_0) \right]$$

$$\mp \sum_{i=1}^{n} \sum_{j=1}^{n} \frac{\partial^2}{\partial x_1^i \partial x_1^j} \left[ w_\pm^{ij}(\overline{x}(t_1), t_1) p(\overline{x}_1, t_1 \mid \overline{x}_0, t_0) \right] = 0$$

Let us repeat that the equations involving the symbols $v_+$ and $w_+$ are related to the direct case where we have $t_1 > t_0$ while the equations involving the symbols $v_-$ and $w_-$ are related to the inverse case where we have $t_1 < t_0$.

Perhaps unexpectedly, the forward and backward versions of the drift velocities and of the diffusion tensors are not completely independent. We have indeed

$$w_+^{ij}(x, t) = w_-^{ij}(x, t) = w^{ij}(x, t)$$

and:

$$v_+^i(x, t) - v_-^i(x, t) = \frac{2}{p(x, t)} \sum_{j=1}^{n} \frac{\partial \left[ p(x, t) w^{ij}(x, t) \right]}{\partial x^j}$$

where we have denoted by $w^{ij}$ the common value of $w_+^{ij}$ and $w_-^{ij}$.

It is very easy to show that the auxiliary vector $u^i$ defined by:

$$u^i = \frac{1}{2}(v_+^i - v_-^i)$$

verifies the relation:

$$p(x, t) u^i(x, t) = \sum_{j=1}^{n} \frac{\partial \left[ p(x, t) w^{ij}(x, t) \right]}{\partial x^j}$$

and that the auxiliary vector $v^i$ defined by:

$$v^i = \frac{1}{2}\left(v_+^i + v_-^i\right)$$

come into play in the conservation equation:

$$\frac{\partial p(x,t)}{\partial t} + \sum_{i=1}^{n} \frac{\partial \left[p(x,t)v^i(x,t)\right]}{\partial x^i} = 0$$

It is natural to call $p(x,t)\bar{v}(x,t)$ the probability current density corresponding to the probability density $p(x,t)$.

A so-called H-theorem proves that the solutions of the Kolmogorov equations always converge to the solution of the Fokker-Planck equation that can thus be considered to represent the ground state of the system.

Some of the results obtained previously are used in the study of stochastic variational principles.

The special case of the stationary processes is considered at length in view of many future applications.

### 3  Non-relativistic stochastic quantum mechanics.

This chapter exhibits the important relations existing between the orthodox quantum theory and the theory of the diffusion Markov processes. We first start from general properties and relations of those processes to deduce the so-called Nelson first equation involving the two vectors u and v and the tensor w. We then apply a stochastic variational principle to a Lagrangian that is the normal counterpart of the classical Lagrangian for a particle with mass m and charge e subjected to an electric potential V and a magnetic potential $\overline{A}$ and this leads to the so-called Nelson second equation involving also the two vectors u and v and the tensor w.

Now Schrödinger equation is transformed in a succession of two changes of variables into a system of two equivalent equations that are identical to the Nelson equations provided that we accept the identity:

$$w^{ij} \equiv \frac{h}{4\pi m} g^{ij}$$

where $g^{ij}$ is the metric tensor in the space of the vectors $\bar{x}$. This procedure unfortunately gives the impression that there exist different pairs of vectors u and v corresponding to different solutions of the Schrödinger equation and particularly to its different particular solutions when the method of separation of variables applies to it (1). This is unpleasant for two reasons at least. First this leads to the objection that the solutions determine the equations, and not the converse, and second it turns out that all pairs of vectors u and v possess poles, except the one corresponding to the ground state. We can however tackle the problem entirely from the side of diffusion Markov processes and Nelson equations. With the proposed form of the tensor w, those latter equations represent a system of two partial differential equations in the two unknown vectors u and v that we can try to solve for a more or less unique solution, by excluding all those that are not sufficiently regular. The selected pair of vectors u and v is then introduced in the Kolmogorov equations and we can try to solve those in their turn, by taking into account the initial condition eventually.

---

[1] One of the first objections against Nelson paper was precisely articulated around this idea. Just to insist (if necessary) on the fact that the theory of stochastic process is a very difficult subject, it can be said that other objections actually proved ignorance in the basic properties of conditional probabilities.

We shall demonstrate the merits of this method by applying it to the harmonic oscillator that is characterized by the potential:

$$V(x) = k x^2 / 2$$

or equivalently by the force:

$$f(x) = -\frac{dV(x)}{dx} = -k x$$

for some positive constant value of k. In this case the Nelson equations for the unknown (one-dimensional) vectors u and v are:

$$\begin{cases} \dfrac{\partial u}{\partial t} + \dfrac{\partial}{\partial x}(u v) + \dfrac{h}{4 \pi m} \dfrac{\partial^2 v}{\partial x^2} = 0 \\ \dfrac{\partial v}{\partial t} + v \dfrac{\partial v}{\partial x} - u \dfrac{\partial u}{\partial x} - \dfrac{h}{4 \pi m} \dfrac{\partial^2 u}{\partial x^2} = -\omega^2 x \end{cases}$$

where we have introduced the auxiliary constant:

$$\omega^2 = k/m$$

With the solutions:

$$\begin{cases} v = 0 \\ u = -\omega x \end{cases}$$

the Kolmogorov equations take the form:

$$\begin{cases} \dfrac{\partial p(x,t|x_0,t_0)}{\partial t_0} - \omega x_0 \dfrac{\partial p(x,t|x_0,t_0)}{\partial x_0} + \dfrac{h}{4 \pi m} \dfrac{\partial^2 p(x,t|x_0,t_0)}{\partial x_0^2} = 0 \\ \dfrac{\partial p(x,t|x_0,t_0)}{\partial t} - \omega \dfrac{\partial [x p(x,t|x_0,t_0)]}{\partial x} - \dfrac{h}{4 \pi m} \dfrac{\partial^2 p(x,t|x_0,t_0)}{\partial x^2} = 0 \end{cases}$$

and if we introduce the new dimensionless variables:

$$\begin{cases} \mathrm{t} = \omega t & \mathrm{x} = \sqrt{\dfrac{2 \pi m \omega}{h}} x \\ \mathrm{t}_0 = \omega t_0 & \mathrm{x}_0 = \sqrt{\dfrac{2 \pi m \omega}{h}} x_0 \end{cases}$$

we can write the Kolmogorov equations in the simpler forms:

$$\begin{cases} \dfrac{\partial p(x,t\mid x_0,t_0)}{\partial t_0} - x_0 \dfrac{\partial p(x,t\mid x_0,t_0)}{\partial x_0} + \dfrac{1}{2}\dfrac{\partial^2 p(x,t\mid x_0,t_0)}{\partial x_0^2} = 0 \\ \dfrac{\partial p(x,t\mid x_0,t_0)}{\partial t} - \dfrac{\partial[x\,p(x,t\mid x_0,t_0)]}{\partial x} - \dfrac{1}{2}\dfrac{\partial^2 p(x,t\mid x_0,t_0)}{\partial x^2} = 0 \end{cases}$$

The method of separation of the variables shows that the first Kolmogorov equation is equivalent to the system of the two equations:

$$\begin{cases} \dot{T}(t_0) - \lambda\, T(t_0) = 0 \\ X''(x_0) - 2x\, X'(x_0) + 2\lambda\, X(x_0) = 0 \end{cases}$$

where the dot symbol denotes the derivation with respect to the time variable $t_0$ and where the prime symbol denotes the derivation with respect to the space variable $x_0$. We know that the Hermite polynomials $H_n(x)$ are the solutions of the equations:

$$H_n''(x_0) - 2x\, H_n'(x_0) + 2n\, H_n(x_0) = 0$$

The particular solutions of the first Kolmogorov equation are thus given by:

$$X^0(x_0) T^0(t_0) = H_n(x_0) \exp(n\, t_0)$$

We see also that the second Kolmogorov equation is equivalent to the system of the two equations:

$$\begin{cases} \dot{T}(t) + \lambda\, T(t) = 0 \\ X''(x) + 2x\, X'(x_0) + 2X(x) + 2\lambda\, X(x) = 0 \end{cases}$$

where the dot symbol denotes the derivation with respect to the time variable t and where the prime symbol denotes the derivation with respect to the space variable x. We can easily verify that the functions:

$$G_n(x) = \exp(-x^2)\, H_n(x)$$

are the solutions of the equations:

$$G_n''(x) + 2x\, G_n'(x) + 2G_n(x) + 2n\, G_n(x) = 0$$

The particular solutions of the second Kolmogorov equation are thus given by:

$$X(x) T(t) = \exp(-x^2) H_n(x) \exp(-n\, t)$$

Taking into account the orthogonality relations verified by the Hermite polynomials and remembering that $H_0(x)$ is equal to 1 we can write:

$$p(x,t) = \frac{1}{\sqrt{\pi}} \exp(-x^2)$$

and:

$$p(x,t\mid x_0,t_0) = \frac{\exp(-x^2)}{\sqrt{\pi}} \sum_{n=0}^{\infty} \frac{H_n(x) H_n(x_0)}{2^n\, n!} \exp[-n(t-t_0)]$$

It is somewhat surprising to see that the solutions of Kolmogorov and Schrödinger equations are expressed with the same set of eigenfunctions and eigenvalues. However a new interpretation of the quantum mechanics based on the conditional probability density may no longer speak about stable states because the exponentials depending on the time appearing in the expression for $p(x,t|x_0,t_0)$ all converge to zero, except the first, when the time t increases indefinitely. Whatever the circumstances are, the harmonic oscillator will always attempt to go or to return to a stable state characterized by:

$$\lim_{t=\infty} p(x,t \mid x_0,t_0) = \frac{1}{\sqrt{\pi}} \exp(-x^2) = p(x)$$

This method has also been applied successfully to the Coulomb field in polar and in parabolic co-ordinates, and before all to the free particle where Schrödinger equation faces serious problems. The remarkable point is that all those cases can be handled by the method of separation of variables and that the particular solutions obtained for the Kolmogorov equations are very close to the particular solutions of the Schrödinger equation. In other words, the vectors u and v corresponding to the particular solution representing the ground state allow anyhow recovering all the other particular solutions, so that the previously raised objection vanishes. There is however an important difference between those solutions because their behaviour with respect to time is no longer periodic in time as dictated by the presence of exponentials such as $\exp(\pm i\, K t)$, but is now severely damped as imposed by the presence of exponentials such as $\exp(-Kt)$. This is actually a confirmation of the H-theorem.

Stationary systems deserve special attention and in the special case where the vector v is identically zero, we show that an expression containing the probability density $p(x)$ is actually a solution of the Schrödinger equation.

Finally, we analyse the two body problem and show how curiously the total and reduced masses introduce in the context of stochastic quantum mechanics.

## 4 Comparison with orthodox theory

This chapter compares orthodox and stochastic quantum theories. This is expected to be the hotly debatable part of this work. It must be insisted on the fact that the stochastic theory does not endanger any of the practical results of quantum mechanics, but that it is at variance with the Copenhagen interpretation. So for example the principles of duality and complementarity are totally unnecessary because matter waves appear absolutely nowhere in this work.

Let us collect hereafter the results that are in favour of the stochastic quantum theory in opposition to the Copenhagen interpretation.

### 4.1 Contraction of the wave packet

According to the orthodox quantum theory, the state of the particle is represented before a measurement by a wave function $\psi_0(\bar{x},t)$, whose gradual changes with time are governed by Schrödinger equation. At the time $t_1$ of the measurement, the state of the particle changes suddenly from whatever value $\psi_0(\bar{x},t)$ it had to a new value $\psi_1(\bar{x},t_1)$. No need to say that such a discontinuous behaviour is not controlled by Schrödinger equation. After the measurement is complete, the state of the particle is represented by the wave function $\psi_1(\bar{x},t)$ whose gradual changes with time are again governed by Schrödinger equation and this will continue until the next measurement. So, we can na-ively say that Schrödinger equation is always at work, except at time $t_1$ where the measurement process imposes to the wave function a new initial condition.

Stochastic quantum theory tells us a similar story in a very different context. According to it, the state of the particle is described by some conditional probability density $p(\bar{x},t|\bar{x}_0,t_0)$ verifying the initial condition $p(\bar{x},t_0|\bar{x}_0,t_0) = \delta(x-x_0)$ and whose gradual changes with time are governed since time $t_0$ and forever by the second Kolmogorov equation. At the time $t_1$ of the measurement, nothing special happens to the conditional probability density $p(\bar{x},t|\bar{x}_0,t_0)$, but we can start using the other conditional probability density $p(\bar{x},t|\bar{x}_1,t_1)$ verifying the presently perfectly well defined new initial condition $p(\bar{x},t_1|\bar{x}_1,t_1) = \delta(x-x_1)$ and whose gradual changes with time will be governed since time $t_1$ and forever by the second Kolmogorov equation. After the measurement, the status of the particle is apparently equally well described by any of the conditional probability densities $p(\bar{x},t|\bar{x}_0,t_0)$ or $p(\bar{x},t|\bar{x}_1,t_1)$ but they do not actually yield equivalent results when used for evaluating the probability to find the particle in a given region of space V. With obvious notations, those probabilities are respectively given by:

$$P_0\{\bar{x} \in V | \bar{x}_0,t_0\} = \int_V p(\bar{x},t|\bar{x}_0,t_0)d\bar{x}$$

and:

$$P_1\{\bar{x} \in V | \bar{x}_1,t_1\} = \int_V p(\bar{x},t|\bar{x}_1,t_1)d\bar{x}$$

Remember however that the conditional probability density $p(\bar{x},t|\bar{x}_1,t_1)$ is by definition subjected to the initial condition $p(\bar{x},t_1|\bar{x}_1,t_1) = \delta(\bar{x}-\bar{x}_1)$ so that the corresponding probabilities will give exact evaluations at $t = t_1$ and sharp estimates for times t only slightly greater than $t_1$. On the contrary, the conditional probability density $p(\bar{x},t|\bar{x}_0,t_0)$ that has already evolved during a much longer time would have flattened and spread over space, leading to fuzzy probabilities at best. We must therefore prefer using the conditional probability density $p(\bar{x},t|\bar{x}_1,t_1)$ at least until the results of the next measurement are available.

But it is perhaps a bad idea to forget the past experience related to time $t_0$ and it is perhaps a good idea to use the more complex conditional probability density $p(\bar{x},t|\bar{x}_1,t_1;x_0,t_0)$ in order to obtain yet better results. Remember however that the stochastic process supporting the stochastic quantum theory is supposed to have the Markov property:

$$p(\bar{x},t|\bar{x}_1,t_1;x_0,t_0) = p(\bar{x},t|\bar{x}_1,t_1)$$

so that only the most recent information is relevant.

### 4.2 Uncertainty relations

We have shown that uncertainty relations exist that simply describe a property inherent to all diffusion Markov processes. They are always at work and are in some sense objec-tive because they do not depend on a measurement being done with or without an observer being present.

In the case of the orthodox quantum theory the uncertainty relations were presented for the first time by Heisenberg who made convincing remarks on the impossibility to have the error $\Delta q$ on the position of a particle and the error $\Delta p$ on its momentum arbitrarily small when measured simultaneously. Physicists rightly formulated the uncertainty principle according to which particles are not the kind of things that can have exact position and exact momentum at the same time. So clearly something must be abandoned and in a kind of grave crisis of despair their opinion shifted progressively from uncertainty of measurement to ideological indeterminacy of existence. This culminated with the idea that because we cannot know what happens between two measurements, it is best to think that nothing happens and that it is meaningless to ask if the position a particle remains the same, it its position is changing with time, if the particle is at rest, if it is in motion, etc...

The role played by the measurements in those deductions left the door open for an obvious objection. Let us perform at time $t_1$ a first precise measurement $q_1$ of the position and let us perform at a later time $t_2$ a second precise measurement $q_2$ of the position. If we invoke the classical variational principle and solve the Euler-Lagrange equations with the boundary conditions:

$$\begin{cases} q(t_1) = q_1 \\ q(t_2) = q_2 \end{cases}$$

we can fill the gap between the two measurements by a definite path with definite positions and momenta at all intermediate instants. The second measurement thus reveals which value $p$ of the momentum the particle had just acquired after the first measurement to bring it from the first point to the second point on a continuous path in the required interval of time.

What is wrong in the previous reasoning independently on the fact that classical mechanics applies or not is that the deduction of the Euler-Lagrange equation from the variational principle implicitly supposes that the solutions possess a continuous derivative. This hypothesis is not fulfilled by the sample functions of diffusion Markov processes that are everywhere continuous, but possess nowhere a derivative with respect to the time. Such functions came as a surprise when mathematicians exhibited the first examples of them a long time ago, but they are now more familiar at the point that we have not hesitated to endow the trajectories of the particles with those curious characteristics. So finally the Heisenberg uncertainty relations in the extreme circumstance where we have $\delta q = 0$ and $\delta p = \infty$ can be used to support our choice of the diffusion Markov processes as the base of our stochastic theory of quantum mechanics.

**4.3  Ehrenfest theorem**

It is proven that some well-known definitions and differential equations of classical mechanics hold good in the stochastic quantum theory provided that we replace the intervening quantities by their mathematical expectations. This is the case for the definition of the velocity and for the differential equations concerning the linear and angular momenta.

## 4.4 Ergodic theorem

Born statistical interpretation of the state function $\psi$ asserts that the predictions of quantum mechanics coincide with the mathematical expectations of well-chosen operators calculated with the probability density $\psi\psi^*$. This has led some physicists to extend this interpretation further by asserting that the formalism of quantum mechanics is applicable only to groups of similar events and should not be applied to isolated events.

This extreme conclusion is absolutely not valid for the stochastic theory. We borrow from classical statistical mechanics the idea that the result obtained when measuring a physical quantity represented by a function $f(\bar{x})$ for a single particle is given by a time average such as:

$$\langle f \rangle = \frac{1}{T} \int_0^T f(\bar{x}(t)) dt$$

calculated along a sufficiently long portion of the trajectory actually followed by that particle. If we transpose this idea in the context of the stochastic theory this time interval must be evaluated along a sufficiently long portion of the sample function that represents the trajectory actually followed by the particle. Of course this is impossible because this particular sample function is completely unknown. In the case of a stationary diffusion Markov process however the so-called ergodic theorem asserts that all those time averages are equal and coincide with the mathematical expectation of the function $f(\bar{x})$ calculated with the probability density $p(\bar{x})$.

## 4.5 Hidden variables

The present theory is based on the properties of a "true" diffusion process, that is one for which that the diffusion tensor is not identically equal to zero. We must then give up as hopeless the idea that the introduction of hidden variables could restore determinism. This would lead indeed to the contradictory conclusion that the original diffusion tensor was identically equal to zero.

## 4.6 Phase space distribution.

It is impossible to replace the probability densities $p(\bar{x},t)$, $p(\bar{x}_0,t_0,\bar{x}_1,t_1)$ and conditional probability density $p(\bar{x}_1,t_1|\bar{x}_0,t_0)$ by a supposedly more precise probability density $p(\bar{x},\bar{v},t)$ which could depend not only of the position $\bar{x}$, but also on the velocity $\bar{v}$ if it exists. This would also lead to a diffusion tensor identically equal to zero.

## 4.7 H-theorem

The H-theorem asserts that the conditional probability density $p(x_1,t_1|x_0,t_0)$ converges to the probability density $p(x_1,t_1)$ when the time difference $t_1 - t_0$ converges to $\infty$. This is illustrated by the fact that in the case of the free particle, linear harmonic oscillator and Coulomb field, the particular solutions of the Kolmogorov equation depend on the time t via exponentials such as $\exp(-Kt)$ while the particular solutions of the Schrödinger equation depend on the time via exponentials such as $\exp(\pm iKt)$. This explains in a very simple way the experimental fact that an excited atom for example returns quite rapidly to its ground state.

# 5 Relativistic stochastic quantum mechanics.

This chapter is clearly unfinished. The definitions and equations for relativistic diffusion Markov processes are extrapolated in a natural manner from their non-relativistic counterparts, but this must be done with much care and not simply formally. Let us remember that the stochastic quantum theory must solve the two Nelson equations for three unknowns. In the non-relativistic domain, their relations with Schrödinger equation lead unambiguously to a well defined value for the tensor w and everything is thus fine. There is no such deus ex machina here. On the one hand, Klein-Gordon equation does not do the trick and this is not a bad thing after all because it does not lead to the correct energy levels for the hydrogen atom, and on the other hand we have been unable to find any such relations with Dirac equation.

# 6 Covariance of stochastic equations

This chapter deals with the covariance of the quantum stochastic equations. It has been developed not only because it was mathematically interesting in itself, but also because it is perhaps the entrance door to the relativistic domain and (why not?) to the long searched theory unifying quantum mechanics and general relativity.

# 7 Conclusions

We hope that the reading of this work will inspire young(er) mathematicians and physicists and invite them to develop it. There are many questions left unanswered. So for example, the spin appears nowhere because we have left aside the magnetic properties of atoms. This could be a corollary of the relativistic version of the stochastic quantum theory, as Dirac results suggest it, but the correct form of the diffusion tensor is not yet known for sure.

Are you hesitating to embark in that direction because you object that stochastic theory would have been obviously proposed in preference to orthodox quantum theory if it was so far better? The answer simply lies in the fact that analytic probability theory has been developed only relatively recently, is a difficult theory and suffers therefore from its esoteric character. Let us remember some historical facts to make things clearer. The most often quoted milestones in the history of orthodox quantum theory are:

1900 = Planck theory of black body radiation
1905 = Einstein explanation of photo-electric effect by means of the photons
1913 = Bohr theory of atomic spectra
1926 = Copenhagen interpretation
1926 = Schrödinger equation
1927 = Heisenberg uncertainty principle
1928 = Dirac equation

By the way, the forgotten ideas that could have changed the story of quantum mechanics are:

1923 = Duane paper
1924 = Epstein-Ehrenfest paper

In contrast, in the case of the modern theories of probability, random variables and stochastic processes, we simply quote:

1931 = Kolmogorov equations for diffusion Markov processes

1933 = Axiomatization of probability by Kolmogorov ([1])
1943 = Riesz ergodic theorem
1968 = Nelson paper

---

[1] Since or because of the grave crisis on the foundations of mathematics due to the many paradoxes inherent to the theory of sets in its primitive form, mathematical texts often follow the example given by the monumental work of Bourbaki and present their theorems, propositions, corollaries, ... as developing logically from well chosen axioms. History however tells us something different about the genesis of theories. So, random variables were manipulated by probabilists long before it was recognized that the mathematical concept involved was that of measurable functions, and in fact long before measure theory was invented. In the completely different domain of arithmetic, more or less intricate operations and algorithms have been used successfully for several thousands of years before Peano proposed his system of axioms for the natural numbers.